# CONCEPTION INDIVIDUELLE ET COLLECTIVE. APPROCHE DE L'ERGONOMIE COGNITIVE
## Willemien Visser

INRIA (Institut National de Recherche en Informatique et en Automatique)

EIFFEL2 (Cognition & Coopération en Conception)

Rocquencourt B.P.105

78153 LE CHESNAY CEDEX (FRANCE)

tél.: 01 39 63 52 09

email:  Willemien.Visser@inria.fr

fax: 01 39 63 59 95

http://www-c.inria.fr/eiffel/index.html

**Résumé :** Ce texte présente l'approche de l'ergonomie cognitive de la conception, dans ses expressions individuelle et collective. Il est centré sur la conception conduite collectivement dans ses rapports avec la conception individuelle. L'accent est sur les aspects cognitifs examinés à travers des études empiriques. Le cadre théorique adopté est celui du traitement de l'information, spécifié pour des problèmes de conception.

On présente les caractéristiques cognitives des problèmes de conception: effets du caractère mal-défini et des différents types de représentation mise en œuvre dans leur résolution de ces problèmes, entre autres sur le caractère plus ou moins "satisfaisant" des différentes solutions possibles. On décrit dans un premier temps les activités cognitives mises en œuvre aussi bien dans la conception conduite individuellement que dans sa forme collective: différents types d'activités de contrôle et d'activités exécutives de développement et d'évaluation de solutions. On introduit ensuite les caractéristiques spécifiques de la conception collective: d'une part, des activités de co-conception et de conception distribuée, de synchronisation temporo-opératoire et de synchronisation cognitive, d'autre part, des spécificités de la conception collective comme différents types d'intervention dans le processus de la part des co-concepteurs, d'évaluation des solutions proposées et d'argumentation. Dans la conclusion, on confronte les deux types de situation de conception, individuelle et collective.

**Mots-clés** : Conception individuelle, Conception collective, Ergonomie cognitive, Psychologie cognitive, Processus cognitifs, Coopération.

# Individual and Collective Design. The Cognitive-Ergonomics Approach

**Abstract**: This text presents the cognitive-ergonomics approach to design, in both its individual and collective form. It focuses on collective design with respect to individual design. The theoretical framework adopted is that of information processing, specified for design problems.

The cognitive characteristics of design problems are presented: the effects of their ill-defined character and of the different types of representation implemented in solving these problems, amongst others the more or less "satisficing" character of the different possible solutions. The text first describes the cognitive activities implemented in both individual and collective design: different types of control activities and of the executive activities of solution development and evaluation. Specific collective-design characteristics are then presented: co-design and distributed-design activities, temporo-operative and cognitive synchronisation, and different types of argumentation, of co-designers' intervention modes in the design process, of solution-proposals evaluation. The paper concludes by a confrontation between the two types of design, individual and collective.

**Keywords**: Individual design, Collective design, Cognitive ergonomics, Cognitive psychology, Cognitive processes, Cooperation.

## Table des matières

## 1. Introduction

Dans le langage courant, pré-scientifique, les termes "conception" et "concepteur" renvoient généralement à un statut, à une fonction sociale. Selon cette acception, la conception consiste en des activités caractérisées par leur nature fortement conceptuelle, qui sont "menées par des professionnels hautement qualifiés qui ne participeront que rarement à la réalisation de l'objet même à concevoir (l'artefact)" (Darses & Falzon, 1996, p. 123)[1]. Un exemple typique en est constitué par les activités de l'architecte qui pense le concept d'un édifice, ou de l'ingénieur du bureau d'étude (B. E.) définissant les spécifications et le concept d'un produit, qui sera à réaliser par des opérateurs[2] de fabrication (id.). Dans l'approche adoptée ici, c'est-à-dire celle de l'ergonomie cognitive, les notions "conception" et "concepteur" sont définies sur d'autres bases et couvrent plus d'activités et de personnes. "Conception" renvoie dans ce cadre à un type d'activité de résolution de problèmes (qui sera détaillé ci-dessous)[3]. Quand on souscrit à cette définition, les activités de conception ne sont plus le monopole des ingénieurs du B.E. Dans ce cas, d'autres opérateurs sont également amenés à avoir des activités qui, dans une analyse en ergonomie cognitive, relèvent de la conception. Il peut s'agir de dessinateurs ou d'autres techniciens du B.E., et d'opérateurs travaillant à l'atelier qui sont censés "seulement" "réaliser" ce qui a été défini conceptuellement par les ingénieurs. Même des sous-traitants chargés de manufacturer des composants, et des usagers peuvent intervenir en tant que concepteurs quand ils contribuent à l'élaboration (établissement ou modification) de la solution, qui, sous forme implémentée, constituera l'objet à concevoir.

Dans ce texte, nous présentons une approche d'ergonomie cognitive de la conception, dans ses expressions individuelle et collective. Notre centre d'intérêt sera la conception conduite collectivement, dans ses rapports avec la conception individuelle. L'accent est sur les aspects cognitifs, examinés à travers des études empiriques. Le cadre théorique adopté est celui du traitement de l'information (Newell & Simon, 1972), spécifié pour des problèmes de conception (Visser, 1991).

<u>Plan de ce texte</u>. Après cette brève introduction, nous commençons par confronter l'approche de l'ergonomie cognitive à d'autres points de vue sur la conception, notamment collective (section 2). Dans la section 3, nous caractérisons brièvement les problèmes de conception d'un point de vue cognitif. Nous décrivons ensuite les différents types d'activité mise en œuvre dans le processus de conception (sections 4 et 5). Une première présentation s'appuie surtout sur des études de la conception conduite individuellement (section 4). Dans la section 5, nous introduisons les caractéristiques spécifiques de la conception collective par rapport à la conception individuelle. Dans la conclusion, nous confrontons les deux types de situation de conception, individuelle et collective.

---

[1] Pour l'élaboration de ce texte, je suis bien sûr redevable aux travaux de mes collègues, notamment à deux articles de Françoise Darses et Pierre Falzon (Darses, 1997; Darses & Falzon, 1996).

[2] En ergonomie, le terme "opérateur" renvoie à "celui qui exécute certaines tâches dans certaines situations de travail bien précises". Dans ces tâches, il y a mise en œuvre d'"une activité orientée vers des objectifs précis, avec des contraintes précises" (De Montmollin, 1997).

[3] Une tâche demande de la "résolution de problèmes" d'une personne si la solution ne peut pas être récupérée telle quelle en mémoire et/ou si la représentation que cette personne se fait de sa tâche ne lui permet pas de récupérer en mémoire une ou des procédures toutes faites lui permettant d'atteindre son but (Hoc, 1987).

## 2. Approche de l'ergonomie cognitive

L'ergonomie vise à améliorer la compatibilité entre les opérateurs et leurs systèmes de travail, informatiques ou autres. L'ergonomie cognitive s'intéresse aux aspects cognitifs des situations de travail (Green & Hoc, 1991).

L'intérêt relativement récent de l'ergonomie pour les situations de conception collective concorde avec l'évolution des situations de conception en milieu industriel, où les enjeux d'assistance au travail collectif sont devenus cruciaux. Un enjeu majeur de la modernisation des entreprises est, en effet, la création de nouvelles organisations de conception qui reconnaissent le caractère collectif du travail et le décloisonnement des différents métiers, notamment ceux de conception et de fabrication.

La spécification de l'assistance aux activités de conception nécessite des modèles cognitifs de ces activités. C'est le seul moyen de s'assurer de la compatibilité entre, d'une part, les processus et représentations des concepteurs et, d'autre part, les caractéristiques des méthodes et autres outils de conception. Anticipant des résultats présentés ci-dessous, nous pouvons donner un exemple. Il s'agit du constat que l'activité de conception est organisée de façon "opportuniste" (pour une définition, v. ci-dessous). Cette conclusion, qui s'appuie sur des données recueillies dans le contexte de projets de conception industrielle, a mis en cause le bien-fondé des modèles utilisés le plus souvent comme base des outils informatiques et autres, c'est-à-dire des modèles qui représentent l'organisation de la conception comme l'implémentation d'une planification hiérarchique[4]. Des exemples d'outils présupposant une telle planification sont, d'une part, de nombreuses méthodologies (méthodes prescriptives) de conception, et d'autre part, la majorité des environnements informatiques de conception (notamment de logiciel) et d'autres systèmes classiques censés offrir une assistance à des tâches de conception (notamment des systèmes CAO) (Visser & Hoc, 1990).

Au plan collectif, le développement des différentes modalités d'assistance à la coopération inter-métiers et intra-métiers requiert en outre des modèles cognitifs des mécanismes de coopération mis en jeu dans les situations de conception collective. Cette assistance peut prendre la forme de méthodes (par exemple, d'ingénierie concourante ou intégrant des logiques de conception) ou d'autres outils (par exemple, de CFAO et d'IAO ou, de manière assez différente, sur la base de systèmes argumentatifs).

En ce qui concerne l'activité de conception collective, l'approche de l'ergonomie cognitive diffère d'autres perspectives en ergonomie. Elle se distingue notamment de l'approche qui analyse cette activité sous l'angle de la conduite de projets de conception dans l'objectif de définir les apports et les modalités de l'intervention ergonomique lors de la conception de situations de travail (Daniellou, 1988), ou de préciser la mise en œuvre de démarches de conception participative (Garrigou, 1992). Le point de vue de l'ergonomie cognitive se distingue aussi d'approches d'autres disciplines devant l'activité de conception collective. Il s'agit, par exemple, d'analyses organisationnelle, sociale, psychosociale ou sociotechnique (Darses, 1997; Darses & Falzon, 1996).

---

[4] La distinction que nous faisons entre "planification" et "organisation" renvoie à la différence entre la construction (et l'utilisation) par un concepteur de représentations mentales dans le dessein d'organiser son activité de conception ("planification") et la structuration de l'activité de conception effective, telle qu'elle peut être observée par un observateur extérieur ("organisation") (Visser, 1994a; Visser, 1994b).

## 3. Caractéristiques cognitives de la conception

D'un point de vue cognitif, la conception peut être caractérisée par des propriétés (i) du problème à résoudre, (ii) des représentations et des processus de résolution mis en œuvre pour le résoudre, et (iii) de la solution développée (pour des revues de question générales, v. Falzon et al., 1990; Visser, 1992a; pour des aspects méthodologiques de recueil de données sur des problèmes de conception et d'autres problèmes "mal définis", v. Visser & Falzon, 1988; Visser & Morais, 1988).

• Un problème de conception est souvent qualifié de problème "mal défini" (Reitman, 1964; Reitman, 1965) ou "mal structuré" (Newell, 1969; Simon, 1973/1984)[5]: il n'est pas spécifié complètement, de façon immuable et non ambiguë par les spécifications initiales, qui souvent énoncent des contraintes conflictuelles et sont exprimées en termes de nature différente de celles de la solution (Lawson, 1979/1984). Les variables d'un problème de conception et leurs relations ne peuvent être scindées en sous-systèmes indépendants.

Une partie importante de l'activité de conception consiste à spécifier le couple problème-solution. Dès le moment que le concepteur[6] commence à construire une représentation des spécifications initiales de problème et qu'il entreprend leur analyse, il commence à spécifier une solution[7]. Ceci se fait notamment en résolvant des conflits entre contraintes.

• Des représentations de différentes natures sont mises en œuvre dans la résolution d'un problème de conception. Elles diffèrent notamment quant au domaine de référence, au niveau d'abstraction et au point de vue[8]. Relativement au premier aspect, les connaissances invoquées sont, par exemple, des méthodes de conception et des connaissances dans le domaine d'application. En ce qui concerne le niveau d'abstraction en conception architecturale, par exemple, (Lebahar, 1983) a constaté que le type de production graphique varie en fonction de l'étape de conception dans laquelle l'architecte est engagé: avant de finir par devenir des plans précis, cotés, ces productions sont floues dans les étapes initiales. Ce caractère flou a une motivation: c'est ainsi qu'une réduction de l'incertitude trop précoce peut être évitée.

• Il n'y a pas de solution unique à un problème de conception. La solution adoptée en finale n'est pas ou bien "correcte", ou bien "incorrecte": elle est une solution acceptable parmi d'autres. Les spécifications "finales" auxquelles arrive un concepteur peuvent être contestées, non pas parce qu'elles seraient incorrectes, mais parce que, suivant des critères différents, une autre solution pourrait être adoptée, par le même concepteur ou par un autre (Bisseret, Figeac-Létang, & Falzon, 1988; Bonnardel, 1991). Ceci présuppose que les critères non plus ne peuvent être ordonnés selon un ordre objectif d'importance ou de validité, unanimement accepté par tous les intervenants sur le projet de conception. La solution est plus ou moins satisfaisante: il n'y a pas de critère "définitif" pour tester toute solution proposée, comme il en existe pour des problèmes "bien définis" (Simon, 1973/1984).

---

[5] V. aussi la définition que (Hayes-Roth, Hayes-Roth, Rosenschein, & Cammarata, 1979) donnent des problèmes de "génération" (que les auteurs opposent aux problèmes d'"interprétation").

[6] Notre utilisation dans ce texte du terme "concepteur" au masculin ne signifie pas que tous les concepteurs soient des hommes.

[7] Pour cette raison, il est plus approprié de parler de "couples problème-solution" que de deux entités séparées —"le" problème et "sa" solution— qui existeraient indépendamment et seraient développées consécutivement. C'est pour des raisons de simplicité que nous continuerons à utiliser dans ce texte les termes "problème" et "solution", selon l'aspect qui est privilégié dans le propos (notamment "analyse d'un problème" ou "développement d'une solution").

[8] Faute de définition plus précise de la notion de "point de vue", celle-ci renvoie ici aux représentations et savoirs propres aux représentants des différents métiers en présence sur un projet de conception, par exemple, en ingénierie, les concepteurs de la structure, les concepteurs du système, les opérateurs de l'atelier et ceux de la maintenance.
Dans une autre optique —pas nécessairement contradictoire avec la précédente—, cette notion renvoie aux différents niveaux d'abstraction sur un axe abstrait-concret, qui se traduit concrètement par des analyses fonctionnelle, structurelle et physique du problème-solution.

## 4. Activités mises en œuvre dans la conception

Globalement, pour résoudre un problème —de conception ou autre—, une personne procède à des activités de trois types: (i) construction d'une représentation mentale de problèmes (ou "formulation de problèmes" en termes d'autres auteurs), (ii) développement (ou "génération") d'une solution, et (iii) évaluation de cette solution. Un aspect de la construction d'une représentation d'un problème consiste à analyser celui-ci en le décomposant en sous-problèmes. Dans le cas de la conception, "le" problème par la décomposition duquel un concepteur commence le travail sur un projet de conception correspond à ce projet de conception dans son ensemble[9]. Jusqu'à ce qu'elle soit formulée au niveau d'implémentation, chaque solution à un problème est à raffiner et constitue dans ce sens un nouveau problème à résoudre. "Problème" et "solution" sont ainsi des notions relatives, deux faces d'une même entité.

Quant à l'évaluation, une solution n'est pas nécessairement développée entièrement avant d'être évaluée. Souvent, une première évaluation d'une solution conduit à un développement renouvelé de celui-ci. Une solution peut être développée en partie, reprise ultérieurement, et évaluée en étapes; plusieurs solutions peuvent être développées avant d'être évaluées (pour un exposé plus détaillé de ces différentes modalités, v. Visser, 1991).

Darses (1997) développe deux autres aspects de l'activité de conception non détaillés ici: la prégnance de la solution (la propension de "penser solution") et l'élaboration de la solution par simultanéité de différents points de vue sur l'objet à concevoir.

### 4.1 Développement d'une solution

Le développement d'une solution (souvent considérée comme "la conception proprement dite") peut se faire de différents modes: nous avons distingué l'évocation d'une solution de l'élaboration d'une solution (pour un exposé plus détaillé de ces processus de raisonnement mis en œuvre dans la conception individuelle, v. Visser, 1991).

Evocation de solutions. Comme nous l'avons remarqué déjà (v. Note 3), si un "problème" est connu tel quel d'un concepteur (de sorte que la solution correspondante peut être évoquée telle quelle en mémoire), il ne s'agit pas d'une tâche de "résolution de problèmes", d'un point de vue de la psychologie cognitive.

Si un concepteur ayant construit une (première) représentation du problème à résoudre (le "problème-cible") juge celui-ci similaire à un problème connu (et résolu dans le passé, le "problème-source"), la représentation mentale du problème-cible peut évoquer dans sa mémoire la "solution-source" correspondante. Le développement d'une "solution-cible" sur une telle base renvoie à ce qui est intitulé, depuis une quinzaine d'années maintenant, "réutilisation"[10] (sous-entendu: de solutions ou de connaissances spécifiques, concernant des problèmes particuliers, c'est-à-dire des problèmes-sources analogues au problème-cible, résolus dans le passé). En effet, de plus en plus d'études montrent que, en conception, à côté de connaissances de haut niveau, c'est-à-dire plutôt abstraites, un rôle important est joué par des connaissances spécifiques, concernant des problèmes particuliers, analogues, résolus dans le passé (Visser, 1999).

---

[9] Nous utilisons, dans la suite de ce texte, les termes de "problème" et de "solution" pour renvoyer aux notions de "problème" ou de "sous-problème", et de "solution" ou de "sous-solution".

Néanmoins, même si on peut réutiliser la solution développée pour un problème analogue, cette solution doit généralement être adaptée dans une mesure plus ou moins importante avant de pouvoir constituer "la" solution au problème posé (v. les études empiriques conduites sur la réutilisation en conception, Visser, 1993b; Visser, 1999). Le mode de raisonnement mis en œuvre dans la réutilisation est le raisonnement analogique (Visser, 1992b; Visser, 1996).

Si le concepteur reconnaît dans le problème-cible une classe de problèmes connue, il pourra procéder par instantiation du schéma de problème correspondant.

<u>Elaboration de solutions</u>. Si les différents modes précités ne sont pas possibles, le concepteur devra élaborer une solution à partir de ses connaissances générales (angl. *reasoning from first principles*). Il s'agit de connaissances plutôt abstraites, c'est-à-dire non spécifiques à des problèmes particuliers, mais portant sur des classes de problème dans les différents domaines concernés par le projet de conception. Pour l'exploitation de ces connaissances, le concepteur peut faire appel aux différentes formes de raisonnement classiques, comme le raisonnement déductif, et à des stratégies et heuristiques fortes et faibles de résolution de problèmes, comme des stratégies de type "fins et moyens" ou "essais et erreurs". Un développement de solutions qui repose exclusivement sur des connaissances générales est qualifié de raisonnement *ex nihilo* (*from scratch*). Le problème correspondant au projet global de conception ne sera jamais résolu *ex nihilo* ; ce mode de résolution ne s'applique qu'à des sous-problèmes. Selon la nature d'un projet de conception et, par conséquent, celle de ses sous-problèmes, la proportion de problèmes qui sera résolue *ex nihilo*, comparée à celle qui le sera par réutilisation, sera plus ou moins importante. Un facteur dont on peut s'attendre à ce qu'il intervient est la dimension routinière du projet de conception, c'est-à-dire du problème global (Brown & Chandrasekaran, 1989; Mayer, 1989; Navinchandra, 1991). Nous ne disposons pas d'études empiriques sur ce point. (Visser, 1992b; Visser, 1996) a examiné le lien entre raisonnement analogique et conception créative à travers l'analyse de trois études empiriques ayant porté sur des projets de conception qui se distinguaient quant à la proportion de tâches routinières. Elle a avancé des arguments pour affirmer que le raisonnement analogique utilisé pour résoudre un problème de conception est susceptible de conduire à des solutions de conception créatives.

## 4.2 Évaluation d'une solution

---

[10] Ou "raisonnement à partir de cas", d'après l'appellation utilisée en Intelligence Artificielle (Visser, 1993b; Visser, 1993c).

L'évaluation d'une solution se fait par confrontation à des critères ou contraintes[11] de différents types (Bonnardel, 1992; v. aussi Darses, 1994).

L'évaluation ne concerne pas seulement les solutions de conception. Les contraintes elles-mêmes sont aussi soumises à évaluation, notamment pour déterminer leur importance respective et les priorités entre elles. Si la première fonction de l'évaluation de solutions est de conduire le concepteur à émettre des jugements d'acceptabilité concernant celles-ci (Bonnardel, 1992), l'évaluation a aussi une fonction de contrôle: elle sert également à orienter l'activité de résolution de problème.

### 4.3 Contrôle de l'activité

A côté des activités "exécutives" de développement et d'évaluation de solutions, le concepteur met en œuvre des activités de "contrôle" (ou de "gestion") de son activité.

L'évaluation, dans sa fonction de contrôle, contribue à déterminer le "focus de l'attention": elle aide le concepteur à décider sur quel aspect du projet de conception il va porter son attention par la suite (par exemple, quel sera le sous-problème ou le pas de raisonnement suivant). En effet, lors de l'évaluation d'une solution, l'attention du concepteur est dirigée vers les aspects positifs ou négatifs de celle-ci. L'évaluation négative d'une solution amène le concepteur à rechercher des solutions permettant de modifier celle-ci, c'est-à-dire de développer une contre-proposition de solution[12], tandis qu'une évaluation positive le conduit à poursuivre sa résolution de problèmes. Cette continuation peut se faire en raffinant la solution qui vient d'être évaluée, ou en développant des éléments de solution pour d'autres sous-problèmes du problème global.

L'organisation par le concepteur de ses activités exécutives a une fonction de contrôle importante. Sur ce point, un résultat qui se dégage de différentes études empiriques est que l'activité de conception est organisée de façon "opportuniste" (Bisseret et al., 1988; Visser, 1987; Visser, 1994b). Suivant en cela Hayes-Roth et Hayes-Roth (1979), ce mode d'organisation de l'activité a été modélisé par les auteurs cités au moyen d'un modèle "tableau noir" (*blackboard*) (pour une présentation générale du modèle, v. Nii, 1986a; Nii, 1986b). L'organisation d'une activité est qualifiée d'"opportuniste" quand le concepteur ne suit pas strictement, c'est-à-dire sans déviations, un chemin prédéterminé selon un plan préétabli[13], mais tire profit de possibilités d'action qui se présentent au cours de son activité et qu'il interprète comme des "opportunités" du point de vue de leur "coût cognitif" (Visser, 1994a; Visser, 1995). En termes de démarches "descendante" et "ascendante" dans une représentation de l'état de la solution en niveaux d'abstraction, il en résulte une suite de mouvements. Ces mouvements se font en descendant plus ou moins en profondeur, pour remonter ensuite plus ou moins vers la surface, et pour redescendre éventuellement.

---

[11] Les termes "critères" et "contraintes" sont utilisés souvent comme synonymes. On peut souhaiter réserver "contraintes" pour des éléments ayant une fonction générative, et "critères" pour ces éléments dans leur utilisation critique, sélective (Visser, 1996). (Darses, 1994) a étudié plutôt le premier type, (Bonnardel, 1992) le second type d'élément.

[12] Dans des textes antérieurs, nous avons utilisé le terme franglais de "solution alternative".

[13] L'exemple d'un tel plan serait de traiter les différents niveaux ou différents points de vue dans un ordre particulier, préétabli (par exemple, de haut-en-bas et en largeur d'abord, ou d'abord le point de vue conceptuel, puis le fonctionnel, et finalement le physique).

En montant entre les niveaux n et n+2, le niveau n+1 n'est pas nécessairement traversé; en descendant de n vers n-2, le niveau n-1 peut être "sauté" (Visser & Hoc, 1990).

Un deuxième résultat intéressant concernant le contrôle de l'activité de conception porte sur le rôle possible de l'analogie à ce niveau. (Visser, 1992b; Visser, 1996) a distingué deux types d'utilisation d'analogies. Au niveau de l'exécution de l'activité (développement et évaluation de solution), le raisonnement analogique est utilisé pour résoudre les problèmes de conception courants (il s'agit de l'utilisation "classique" du raisonnement analogique à laquelle nous avons déjà fait référence plus haut). Au niveau du contrôle de l'activité, ce mode de raisonnement est exploité pour rendre le coût cognitif de l'exécution des actions le plus bas possible. Le premier type est susceptible de conduire à des solutions de conception créatives (v. ci-dessus), tandis que le second type introduit homogénéité et répétition dans un projet de conception, aussi bien en ce qui concerne l'activité que le résultat de celle-ci.

## 5. Conception collective

La caractérisation de la conception que nous venons de présenter jusqu'ici repose presque exclusivement sur des études de la conception conduite individuellement. Si des études expérimentales conduites en laboratoire ou dans d'autres conditions artificiellement restreintes peuvent placer des concepteurs dans une situation de conception exclusivement individuelle, il en va différemment en situation de travail professionnel, où plusieurs participants coopèrent en général sur un projet de conception. Jusqu'à récemment, cependant, les études empiriques dans le domaine ne concernaient pratiquement que la conception individuelle. Ceci vaut également pour d'autres types d'activité. En règle générale, les recherches sur le travail collectif sont récentes (Pavard, 1994; Six & Vaxevanoglou, 1993; Terssac & Friedberg, 1996); v. aussi les travaux dans les domaines du "Computer Supported Cooperative Work" (CSCW) (Bowers & Benford, 1991; Greif, 1988) et de la "cognition distribuée" (résolution de problèmes distribuée, prise de décisions distribuée, Rasmussen, Brehmer, & Leplat, 1991; v. aussi l'Intelligence Artificielle Distribuée).

Pourtant, dès les premières études empiriques sur la conception, quelques auteurs ont étudié la coopération. (Malhotra, Thomas, Carroll, & Miller, 1980), par exemple, ont analysé l'interaction entre concepteurs et clients. Celle-ci a aussi été examinée par (Walz, Elam, Krasner, & Curtis, 1987), qui ont observé par ailleurs la coopération entre les concepteurs eux-mêmes. Les travaux, d'inspiration I.A. notamment, sur la résolution de conflits entre concepteurs travaillant en équipe ont également commencé dès les années 80 (Klein & Lu, 1989). C'est cependant dans les années 90 que les études sur la conception collective ont pris leur envol.

Les études conduites sur la conception individuelle ont fait souvent appel aux techniques de verbalisation provoquée (verbalisation simultanée ou consécutive). Même si le caractère approprié de cette méthode pour l'étude de la résolution de problèmes de conception a été étayé par des arguments probants (Ericsson et Simon, 1984/1993, notamment dans l'introduction à la nouvelle édition de leur ouvrage de référence de 1984), elle introduit bien sûr un artifice dans la situation étudiée. Ceci ne vaut pas pour des situations de conception collective, où les interactions verbales se font sans que l'expérimentateur ait à les provoquer.

Quant à ces interactions, il est utile de dissiper un malentendu possible. Elles ne constituent nullement (exclusivement) une trace d'une soi-disant "véritable" activité de conception qui serait une activité mentale individuelle, que ces interactions verbales permettraient de partager: c'est à travers elles qu'une partie importante de l'activité de conception se fait.

## 5.1 Co-conception et conception distribuée

C'est en faisant référence aux notions de "co-conception" et de "conception distribuée" (Darses & Falzon, 1996) qu'on peut préciser la nature de la coopération de différents concepteurs dans un projet de conception collective.

Co-conception. C'est la situation dans laquelle les concepteurs travaillent conjointement sur le projet de conception. Ils partagent un but commun identique, à l'atteinte duquel chacun contribue selon ses compétences spécifiques. Il s'agit d'une forme de coopération "forte". Quelques études dans lesquelles elle a été examinée sont (Karsenty, 1994, conception d'un modèle conceptuel de base de données), (Darses & Falzon, 1996; Darses, Falzon, & Robert, 1993; Falzon & Darses, 1992, conception de réseaux informatiques), (Bonnardel, 1992; Visser, 1991; Visser, 1993a, conception d'une antenne déployable), (Malhotra et al., 1980, tâches diverses de conception), (Klein & Lu, 1989, conception architecturale) et (D'Astous, Détienne, Robillard, & Visser, 1998; Détienne, Visser, d'Astous, & Robillard, 1999; Robillard, D'Astous, Détienne, & Visser, 1998, inspection technique de logiciel industriel).

Conception distribuée. Dans cette situation, les concepteurs travaillent simultanément, non conjointement, mais en parallèle, sur un projet de conception. Chacun accomplit une des différentes tâches dans lesquelles le projet a été décomposé préalablement et qui lui a été allouée. Chacun a ses propres sous-buts, tout en connaissant le but commun final, et il y contribue indirectement. Cette forme "faible" de coopération a été étudiée par Béguin (1994, dans Darses & Falzon, 1996). Cet auteur a réalisé une étude auprès de concepteurs utilisateurs de systèmes de CAO dans une entreprise d'ingénierie. Il montre que le graphisme, en dessin technique, est utilisé comme outil d'une part de communication et d'autre part de transfert de représentations entre collègues.

Dans des processus d'ingénierie concourante (ou "ingénierie simultanée") (Darses, 1997), on trouve typiquement les deux formes de conception collective. Des phases de co-conception et de conception distribuée y alternent en effet dans une succession d'étapes, chacune bien organisée d'un point de vue méthodologique –même si, en réalité, ce n'est pas toujours le cas. Dans des situations plus classiques, les deux formes sont présentes également, mais l'alternance et l'organisation de chacune des phases est, en général, moins bien structurée.

## 5.2 Synchronisation temporo-opératoire et synchronisation cognitive

L'activité collective s'exprime au travers d'interactions entre les concepteurs guidées par deux objectifs complémentaires (Darses & Falzon, 1996, p. 125): se synchroniser dans le temps et sur le plan de l'action, et se synchroniser sur le plan cognitif.

Synchronisation temporo-opératoire. A travers des activités de coordination, cette synchronisation remplit une fonction de caractère plutôt opératoire —l'allocation des tâches— et une fonction de caractère plutôt temporel —l'articulation des actions à réaliser (leurs déclenchement, séquencement, arrêt, simultanéité et rythme).

L'analyse des observations faites par Béguin (1994, cité ci-dessus) a montré que cette forme de synchronisation est un mécanisme crucial en conception distribuée (Falzon, Darses, & Béguin, 1996).

<u>Synchronisation cognitive.</u> A travers des activités de communication, cette synchronisation a comme objectif "d'établir un contexte de connaissances mutuelles, de construire un référentiel opératif commun (De Terssac & Chabaud, 1990)" (id.). Il s'agit de s'assurer que tous les concepteurs participant à un projet de conception ont, d'une part, connaissance des faits relatifs à l'état de ce projet et, d'autre part, un savoir partagé quant aux connaissances invoquées. Ces faits relatifs à l'état d'un projet sont notamment les données du problème, l'état de la solution, les hypothèses rejetées et celles qui ont été adoptées, et les justifications concernant ces décisions; les connaissances invoquées concernent surtout le domaine d'application et les méthodes de conception utilisées.

L'importance de cette forme de synchronisation a été constatée, par exemple, dans des analyses de réunions d'inspection technique (D'Astous et al., 1998; Détienne et al., 1999; Robillard et al., 1998). Nous avons montré que, à travers des activités de présentation et de demande d'informations, et de formulation d'hypothèses, la synchronisation cognitive occupe quelque 41 % du temps et des activités dans ces réunions qui font l'objet d'étapes ayant lieu à différents moments dans un projet de conception industrielle. Souvent cette synchronisation constitue un prérequis à l'évaluation des solutions, qui est l'objectif déclaré de l'inspection, c'est-à-dire de la tâche prescrite des concepteurs.

### 5.3 De la conception individuelle à la conception collective

Les aspects coopératifs de la conception collective demandent des activités cognitives spécifiques, notamment de coordination, de communication, de synchronisation et de résolution de conflits. On y retrouve, cependant, également les activités identifiées dans les études conduites sur la conception individuelle. Il s'agit aussi bien de celles, exécutives, de développement et d'évaluation de solutions que de celles, gestionnaires, de planification et d'organisation de ces premières activités.

### 5.3.1 Intervention "spontanée" ou "à la demande"

Une première spécificité de la conception collective se situe au niveau des interventions des différents concepteurs participants (c'est-à-dire, leurs contributions individuelles aux interactions). Ces interventions peuvent être "spontanées", ou elles peuvent se faire "à la demande" d'autres participants. Il arrive souvent, par exemple, dans une situation de co-conception, que des problèmes identifiés et énoncés par un concepteur soient résolus par des collègues. Nous avons observé également que des concepteurs faisaient suivre leurs propositions de solutions par des justifications —c'est-à-dire, des présentations explicites d'arguments et/ou des explicitations des critères de choix utilisés pour ces solutions (Darses & Sauvagnac, 1997)— sans aucune demande —explicite— de l'un des collègues. Nous avons constaté aussi que des propositions de solutions par un concepteur soient suivies aussitôt de contre-propositions de solutions par un collègue —sans qu'une évaluation négative ait forcément été exprimée auparavant par l'un des participants (pour d'autres exemples, v. Visser, 1993a).

Ce premier aspect intéressant des situations de co-conception —le caractère souvent spontané, sans requête explicite préalable, des interventions— peut recevoir un début d'explication quand on considère les interventions dans ces situations comme des invitations implicites à réagir: une réaction n'a pas besoin d'être sollicitée explicitement.

### 5.3.2 Caractère explicite ou implicite de l'évaluation

(Falzon & Darses, 1992), dans une étude de la co-conception de réseaux informatiques par un concepteur expert et un concepteur novice, ont observé que, dans la situation examinée, l'évaluation se faisait majoritairement de façon spontanée. Les requêtes —explicites— d'évaluation étaient rares, mais l'évaluation de propositions de solutions était fréquente.

Dans notre étude sur la conception collective d'antenne (Visser, 1993a), il y avait deux cas de figure. Comme noté ci-dessus, nous avons observé que, suite à la proposition d'une solution par un concepteur, des contre-propositions de solutions soient formulées par des collègues, sans que personne n'ait exprimé d'évaluation négative de la première solution. Nous avions interprété cette formulation "immédiate" de contre-propositions de solutions comme traduisant une évaluation négative, implicite de la solution proposée précédemment.

Trois concepteurs travaillaient couramment ensemble sur ce projet d'antenne. A peu près une fois par mois, se tenaient des réunions plus larges, lors desquelles cette petite équipe rencontrait d'autres collègues intervenant sur le projet. Lors de ces réunions, les solution développées par les trois concepteurs étaient évaluées par leurs collègues, experts dans d'autres domaines, donc utilisant d'autres critères —ou les mêmes critères, mais utilisés différemment. L'appel à ces experts pouvait se faire explicitement ou rester implicite.

L'autre cas de figure est celui de l'intervention d'un expert inter-domaine, considéré par ses pairs comme le "super-expert" de l'entreprise. Chacun à leur tour, des concepteurs travaillant sur différents projets faisaient appel à lui lorsqu'ils étaient confrontés à des problèmes particulièrement ardus. Les requêtes d'évaluation adressées à ce "super-expert" étaient plutôt exprimées explicitement.

Ces constats faits dans des études sur des situations de conception dans lesquelles le développement et l'évaluation de solutions ne faisaient pas l'objet d'étapes distinctes peut être rapproché de ce que nous avons observé dans les réunions d'inspection analysées par D'Astous, Détienne, Robillard et Visser. Comme d'autres méthodologies de conception, celle qui était utilisée dans ce projet prescrivait de procéder aux activités de développement et d'évaluation dans des étapes distinctes du projet. Nous avons constaté cependant que, malgré les prescriptions ordonnant explicitement de procéder, pendant les réunions d'inspection, seulement à des activités d'évaluation et de reporter la "conception" à une étape ultérieure, les concepteurs y procédaient également au développement de contre-propositions de solutions. Ces activités de "conception" occupaient même quelque 21 % du temps et des activités pendant les réunions. Selon (D'Astous et al., 1998), ce développement de solutions avait souvent comme fonction l'explicitation d'arguments justifiant l'évaluation —négative— de la solution pour laquelle la contre-proposition était développée.

Evaluation implicite: caractère positif ou négatif. L'expression implicite d'une évaluation prend un aspect différent selon que celle-ci est positive ou négative. Si, après la proposition d'une solution par un concepteur, ses collègues "passent à la suite", on peut supposer que l'évaluation de cette solution était positive. Si la proposition d'une solution est suivie d'une contre-proposition —c'est-à-dire, que l'"on reste sur le même problème"—, on peut conclure à une évaluation négative de la solution.

### 5.3.3 Mouvements argumentatifs

A partir des données recueillies sur l'inspection technique, nous avons effectué une analyse en termes de "mouvements argumentatifs" en jeu dans la conception (Détienne et al., 1999). Pour cela, nous avons cherché à identifier des articulations fréquentes d'activités. Une de ces articulations fréquentes était composée d'une séquence d'activités de trois types, de la synchronisation cognitive, de l'évaluation d'une solution et de la contre-proposition d'une autre solution. Les activités de proposition d'une autre solution pouvaient précéder ou suivre les activités d'évaluation. Nous avons traduit cette articulation en termes du triplet de mouvements argumentatifs PROPOSITION-OPINION-ARGUMENTS. La synchronisation cognitive constitue un prérequis à l'évaluation et en fixe l'objet (PROPOSITION), de sorte que les concepteurs disposent d'une base commune leur permettant de procéder à une évaluation "armée" de la solution (OPINION). Cette évaluation peut demander cependant une justification (ARGUMENTS), à plus forte raison si elle est négative: la contre-proposition d'une autre solution sert alors ce but.

## 6. Confrontation des situations de conception individuelle et de conception collective: conclusion

Notre présentation, dans ce texte, des études empiriques conduites sur la conception a été faite dans un enchaînement allant de l'individuel vers le collectif. Il s'est agi là non seulement d'une succession dans la chronologie des études. Nous estimons que, dans l'état actuel de nos connaissances sur les aspects cognitifs de la conception, il convient de considérer les relations entre activités mises en œuvre dans la conception individuelle et dans la conception collective de la façon suivante. Les activités dans la conception collective sont, d'une part, celles qui sont déployées dans la conception individuelle **plus** un certain nombre d'autres activités spécifiques à la coopération (notamment de coordination, communication, synchronisation et résolution de conflits). D'autre part, ces activités sont organisées de façon plus complexe (v. 5.1 à 5.3). Nous n'avons pas de raisons de supposer que la coopération en conception modifie la nature des processus **élémentaires** de résolution de problème (de développement et d'évaluation de solutions). Nous avons constaté cependant qu'elle modifie l'évocation et l'organisation de ces processus élémentaires —et on peut faire l'hypothèse que les solutions résultantes sont différentes (plus diversifiées et peut-être même "meilleures"). Des exemples de l'organisation spécifique à la conception collective présentés dans ce texte sont les différents modes d'articulation des activités de développement et d'évaluation, selon qu'elles se font de façon spontanée ou à la demande, selon que l'évaluation se fait explicitement ou reste implicite, et selon que ces activités s'articulent dans une suite de mouvements argumentatifs.

Si nous nous sommes centrée sur la conception collective sous la forme de co-conception, c'est surtout faute d'études empiriques sur la conception distribuée (mais v. Béguin, 1994, dans Darses & Falzon, 1996; v. aussi Falzon et al., 1996). Darses et Falzon affirment que les résultats de caractère encore "exploratoire" obtenus dans leurs études et celles de quelques autres collègues tendent à montrer que, "tandis que les situations de co-conception induisent un plus grand recours à une synchronisation cognitive, la conception distribuée met l'accent sur la synchronisation opératoire de la tâche" (Darses & Falzon, 1996, p. 133). Des études récentes sur l'ingénierie concourante (par exemple Martin, Détienne, & Lavigne, 1999) pourront apporter plus de données sur la conception distribuée, et ainsi permettre de continuer la confrontation, non seulement entre situations de conception individuelle et de conception collective, mais également entre différentes situations de conception collective.

## 7. Bibliographie

Daniellou, F. (1988). Ergonomie et démarche de conception dans les industries de processus continu: quelques étapes-clés. *Le Travail Humain, 51*(2), 185-193.

Darses, F. (1994). *Gestion des contraintes dans la résolution des problèmes de conception.* Unpublished Thèse de Doctorat, spécialité Psychologie Cognitive, Université Paris VIII, Saint-Denis.

Darses, F. (1997). L'ingénierie concourante: Un modèle en meilleure adéquation avec les processus cognitifs en conception. In P. Brossard, C. Chanchevrier, & P. Leclair (Eds.), *Ingénierie Concourante. De la technique au social*. Paris: Economica.

Darses, F., & Falzon, P. (1996). La conception collective: une approche de l'ergonomie cognitive. In G. De Terssac & E. Friedberg (Eds.), *Coopération et Conception*. Toulouse: Octarès.

Darses, F., Falzon, P., & Robert, J.-M. (1993, August 8-13). *Cooperating partners: investigating natural assistance.* Paper presented at the HCI'93, Orlando (U.S.A.).

Darses, F., & Sauvagnac, C. (1997). *Conception continue du système de production: une situation pour la construction collective des savoirs techniques*. Paris: CNAM.

De Montmollin, M. (1997). Vocabulaire de l'ergonomie. 2ème édition revue et augmentée (Première édition 1995). Toulouse: Octares.

Détienne, F., Visser, W., d'Astous, P., & Robillard, P. (1999). *Two complementary approaches in the analysis of design team work: the functional and the interactional approach.* Paper presented at the CHI99 Basic Research Symposium, Pittsburgh.

Ericsson, K. A., & Simon, H. A. (1984/1993). *Protocol analysis. Verbal reports as data* (revised edition). Cambridge, Mass.: MIT Press. (First ed. 1984)

Falzon, P., Bisseret, A., Bonnardel, N., Darses, F., Détienne, F., & Visser, W. (1990, 17-19 octobre 1990). *Les activités de conception: L'approche de l'ergonomie cognitive.* Paper presented at the Colloque "Recherches sur le Design", Compiègne.

Falzon, P., & Darses, F. (1992, 23-25 septembre 1992). *Les processus de coopération dans des dialogues d'assistance.* Paper presented at the Congrès de la SELF 92, Lille (France).

Falzon, P., Darses, F., & Béguin, P. (1996, June 12-14). *Collective design processes.* Paper presented at the COOP'96, Second International Conference on the Design of Cooperative Systems, Juan les Pins (France).

Garrigou, A. (1992). *Les apports des confrontations, les orientations socio-cognitives au sein de processus de conception participatifs: le rôle de l'ergonomie.* Thèse de Doctorat, Spécialité Ergonomie, CNAM, Paris.

Green, T. R. G., & Hoc, J.-M. (1991). What is Cognitive Ergonomics? *Le Travail Humain, 54*(4), 291-304.

Greif, I. (Ed.). (1988). *Computer-supported cooperative work: a book of readings*. (Vol. 1). San Mateo, CA: Morgan Kaufmann.
15Daniellou, F. (1988). Ergonomie et démarche de conception dans les industries de processus continu: quelques étapes-clés. *Le Travail Humain, 51*(2), 185-193.

Darses, F. (1994). *Gestion des contraintes dans la résolution des problèmes de conception.* Unpublished Thèse de Doctorat, spécialité Psychologie Cognitive, Université Paris VIII, Saint-Denis.

Darses, F. (1997). L'ingénierie concourante: Un modèle en meilleure adéquation avec les processus cognitifs en conception. In P. Brossard, C. Chanchevrier, & P. Leclair (Eds.), *Ingénierie Concourante. De la technique au social*. Paris: Economica.

Darses, F., & Falzon, P. (1996). La conception collective: une approche de l'ergonomie cognitive. In G. De Terssac & E. Friedberg (Eds.), *Coopération et Conception*. Toulouse: Octarès.

Darses, F., Falzon, P., & Robert, J.-M. (1993, August 8-13). *Cooperating partners: investigating natural assistance.* Paper presented at the HCI'93, Orlando (U.S.A.).

Darses, F., & Sauvagnac, C. (1997). *Conception continue du système de production: une situation pour la construction collective des savoirs techniques*. Paris: CNAM.

De Montmollin, M. (1997). Vocabulaire de l'ergonomie. 2ème édition revue et augmentée (Première édition 1995). Toulouse: Octares.

Détienne, F., Visser, W., d'Astous, P., & Robillard, P. (1999). *Two complementary approaches in the analysis of design team work: the functional and the interactional approach.* Paper presented at the CHI99 Basic Research Symposium, Pittsburgh.

Ericsson, K. A., & Simon, H. A. (1984/1993). *Protocol analysis. Verbal reports as data* (revised edition). Cambridge, Mass.: MIT Press. (First ed. 1984)

Falzon, P., Bisseret, A., Bonnardel, N., Darses, F., Détienne, F., & Visser, W. (1990, 17-19 octobre 1990). *Les activités de conception: L'approche de l'ergonomie cognitive.* Paper presented at the Colloque "Recherches sur le Design", Compiègne.

Falzon, P., & Darses, F. (1992, 23-25 septembre 1992). *Les processus de coopération dans des dialogues d'assistance.* Paper presented at the Congrès de la SELF 92, Lille (France).

Falzon, P., Darses, F., & Béguin, P. (1996, June 12-14). *Collective design processes.* Paper presented at the COOP'96, Second International Conference on the Design of Cooperative Systems, Juan les Pins (France).

Garrigou, A. (1992). *Les apports des confrontations, les orientations socio-cognitives au sein de processus de conception participatifs: le rôle de l'ergonomie.* Thèse de Doctorat, Spécialité Ergonomie, CNAM, Paris.

Green, T. R. G., & Hoc, J.-M. (1991). What is Cognitive Ergonomics? *Le Travail Humain, 54*(4), 291-304.

Greif, I. (Ed.). (1988). *Computer-supported cooperative work: a book of readings*. (Vol. 1). San Mateo, CA: Morgan Kaufmann.
15